
\input phyzzx.tex
\def\a{\alpha}
\def\b{\beta}

\def\d{\delta}
\def\D{\Delta}

\def\r{\rho}

\def\k{\kappa}
\def\l{\lambda}

\def\s{\sigma}

\def\S{\Sigma}

\def\pa{\partial}

\def\ov{\overline}
%
\def\ap#1{{\it Ann. Phys.} {\bf #1}}

\def\pl#1{{\it Phys. Lett.} {\bf B#1}}

\def\prd#1{{\it Phys. Rev.} {\bf D#1}}

\def\np#1{{\it Nucl. Phys.} {\bf B#1}}

%
\REF\call{C.G. Callan, S.B. Giddings, J.A. Harvey, and A. Strominger,
\prd{45}  R1005 (1992).}
\REF\sda{S. P. de Alwis, \pl{289}  278 (1992),\hfill\break
 and \pl{300} 330 (1993).}
 \REF\bc{A. Bilal and C. Callan, Princeton preprint
PUPT-1320,\hfill\break hepth@xxx/9205089 (1992).}
\REF\gs{S. Giddings, and A. Strominger, Santa Barbara preprint
UCSBTH-92-28, hepth@xxx/9207034 (1992).}
\REF\shda{S. P. de Alwis, \prd{46} 5429 (1992)}
\REF\ps{Y. Park and A. Strominger, Santa Barbara preprint
UCSBTH-92-39,
hep-th@xxx/921017 (1992) A. Bilal, Princeton University preprint
PUPT-1373,
hep-th@xxx/9301021 (1993)}
\REF\rt{T. Regge and C. Teitelboim, \ap {88}  286 (1974).}
\REF\ew{E. Witten, \prd{44}  314 (1991)}
\REF\as{A. Strominger \prd{46}, 4396 (1992)}
\REF\rst{J. Russo, L. Susskind, and L. Thorlacius, \prd{46}
3444(1992);\prd{47}
(1993) 533.}
\REF\dea{S. P. de Alwis Colorado preprint COLO-HEP-318,
hep-th/9307140}
\REF\banks{T. Banks, A. Dabholkar, M.R. Douglas, and M O'Loughlin,
\prd{45},
3607 (1992).}
\REF\rrst{J.G. Russo, L. Susskind, L. Thorlacius, \pl{292}, 13
(1992); L.
Susskind and L. Thorlacius,\np{382}, 123 (1992).}.
\REF\bk{A. Bilal and I.I. Kogan, Princeton University preprint,
PUPT-1379,
hep-th@xxx/9301119, (1993)}
\pubnum {COLO-HEP-309\cr hepth/9302144}
\date={February 1993, \cr Revised August 1993}
\titlepage
\vglue .2in
\centerline{\bf Two Dimensional Quantum Dilaton Gravity and the
Positivity of
Energy }
\author{ S.P. de Alwis\foot{dealwis@gopika.colorado.edu}}
\address{Dept. of Physics, Box 390,\break
University of Colorado,\break Boulder, CO
80309}
\vglue .2in
\centerline{\caps ABSTRACT}
  Using an argument due to Regge and Teitelboim, an expression for
the ADM
mass of  2d quantum dilaton gravity is obtained. By evaluating this
expression
we  establish that the quantum theories which can be written as a
Liouville-like theory,  have a lower bound to energy, provided there
is no
critical boundary. This fact is then reconciled with the observation
made
earlier that the Hawking radiation does not appear to stop. The
physical
picture  that emerges is that of a black hole in a bath of quantum
radiation.
We also evaluate the ADM mass for the models with RST boundary
conditions and
find that negative values are allowed. The Bondi mass of these models
goes to
zero for large retarded times, but becomes negative at intermediate
times in a
manner that is consistent with the thunderpop of RST.
\endpage

 In reference [\call](CGHS), a theory of dilaton gravity coupled to
matter was
proposed. Subsequently it was argued in [\sda, \bc ] that the quantum
version
of the theory  must be a conformal
field theory (CFT), and furthermore that it can be transformed into a
solvable
Liouville-like theory.
In this note we address the question of the positivity of energy in
this
theory.   It has been claimed [\gs ] that these theories are sick
since they do
not have a lower bound to energy. Since all quantum dilaton gravity
theories
that have been studied so far can be transformed in to the Liouville
-like
theory (because they all have zero field space curvature) it is
important to
check whether this is indeed the case.  Related to this statement
about the
absence of a lower bound to energy is the observation  [\bc, \sda,
\shda ] that
the black hole solution of the quantum theory seems to radiate
forever.

  Recent work by Park and Strominger [\ps ] seems to indicate that
there might
be a resolution of this problem. They established a positivity
theorem using
arguments derived from supersymmetry considerations, for fields
satisfying
certain asymptotic conditions. However the expressions for the ADM
mass given
there do not give well defined answers for the {\it solutions} of the
classical
theory since the asymptotic conditions of the theorem are violated.
Furthermore
the situation for the Liouville-like theory was left unresolved and
it was not
clear why there appeared to be a conflict with previous work [\gs ].

In this work we resolve  these issues. Indeed we are able to do so
directly for
the quantum theory since the complete space of solutions to  the
Liouville-like
theory is known.
We derive an expression for the ADM and Bondi mass using  arguments
given by
Regge and Teitelboim [\rt ].  We show that the positivity theorems
apply to the
mass as defined by this procedure provided there is no critical
boundary in the
space time such as the one in reference [\rst]. Our argument shows
that the
expression used in [\ew ,\call ,\gs ] actually needs to be modified.
Furthermore unless one imposes boundary conditions on some critical
line[\rst],
corresponding to $r=0$ in four  dimensions, there will be a
contribution from
the negative infinite end of the space.
\foot{This expression is  given already in the second and third
papers of [\sda
] but it was evaluated only at one end of the space.}  Classically
when the
matter stress tensor has zero expectation value (giving static
solutions) the
ADM mass is zero, while for the dynamic solutions (i.e. in the
presence of
collapsing matter) the mass is positive if the incoming matter has
positive
energy. In the quantum case, in the no-boundary theory, the static
solutions
again have zero mass. In the dynamical case with collapsing matter,
there is an
infinite contribution from the negative end of the space. i.e. this
theory
describes  black hole collapse in an infinite bath of radiation.  The
Bondi
mass  can  also be defined. Again in the theory without a boundary
the Bondi
mas is infinite (and positive) at any finite retarded time, but at
positive
infinite retarded time,  it becomes equal to the energy of the
collapsed
matter. This is consistent with the fact that the ADM mass in this
model is
infinite. The Hawking radiation in this picture is just  this
infinite bath
which comes to an end at infinite retarded time leaving behind the
original
mass which collapsed. This is the way that the formalism resolves the
positivity of energy with Hawking radiation which lasts for an
infinitely long
time. It does not reduce the energy from $M_0$ to negative infinity
but from
positive infinity to $M_0$. This is perhaps not a situation which can
give any
insight into the four dimensional
physics of black hole evaporation, but within the well defined
formalism of
this theory of two dimensional quantum gravity, with no
phenomenological
boundary, it seems  to be the only interpretation possible.

With the boundary conditions of Russo Susskind and Thorlacius
(RST)[\rst ] on
the other hand,  we find in the quantum theory, that static solutions
with
negative or zero masses are allowed. In the dynamic case the ADM mass
is just
that of the collapsing matter and thus is positive if the latter is
positive.
We also calculate
the Bondi mass with RST boundary conditions. Here we find that at
negative
infinite retarded time, it goes to the mass of the collapsing matter,
in agreement with the ADM mass of the model, decreases with
increasing time,
and  for positive infinite retarded time,  goes to zero. However at
an
intermediate time the energy becomes negative, and is discontinuous
across a
certain null line; i.e. we encounter the thunderpop of RST[\rst]. The
existence
of  negative Bondi energy is of course consistent with our previous
result that
there are negative energy static solutions.

The Liouville-like action for quantum dilaton gravity [\sda, \bc,
\shda ]
is\foot{We will work with $N$, the number of matter fields, very
large, so that
the difference between $N$ and $\k = {24-N\over 6}$ will be ignored.}

 $$S={1\over 4\pi}\int d^2\s[\mp\pa_{+}X\pa_{-}
X\pm\pa_{+}Y\pa_{-}Y+2\l^2e^{\mp\sqrt{2\over N
}(X\mp Y)}]+S^f+S^{b,c},
\eqn\action$$
where $S^f$ is the conformal matter action, and $S^{b,c}$ is the
reparametrization ghost action. The quantum theory is then given by
functionally integrating with the naive, translationally invariant,
measures.
The field variables $X,Y$ are related to the original variables
$\phi$ (the
dilaton)
 and $\rho$ (half the logarithm of the conformal factor) that occur
in the
 CGHS action gauge fixed to the conformal gauge
$(g_{\a\b}=e^{2\rho}\eta_{\a\b})$, by the following relations;
$$Y=\sqrt{2N}[\r+N^{-1}e^{-2\phi}-{12\over N}\int d\phi e^{-2\phi}\ov
h (\phi
)],\eqn\ycdt$$
 $$X=2\sqrt{12\over N}\int d\phi P(\phi ),\eqn\xcdt$$
where
$$P(\phi )=e^{-2\phi}[(1+\ov h)^2-N e^{2\phi}(1+h)]^{1\over
2},\eqn\pee$$
 $N$ being the number of matter fields. In the above, the functions
$h(\phi),~\ov h(\phi)$ parametrize quantum (measure) corrections that
may come
in when transforming to the translationally invariant measure (see
[\sda, \bc,
\shda ] for details).  The statement that the quantum theory has to
be
independent of the fiducial metric (set equal to $\eta$ in the above)
implies
that this gauge fixed theory is a conformal field theory. The above
solution to
this condition was obtained by considering only the leading terms of
the beta
function equations, but it was conjectured in [\sda, \bc ] (because
of its
resemblance to the Liouville theory,) that it is an exact solution to
the
conformal invariance conditions.\foot{This statement has now been
proved in
[\dea].} It should be noted here that this latter statement is
strictly valid
only when $P$ has no zeroes.  When $N$ is greater than 24 this
implies some
restrictions on the possible quantum corrections, but as shown in
[\shda ]
there is a large class which satisfies these conditions.

 Let us  calculate the time translation
generator by following the argument of Regge and Teitelboim [\rt ].
This goes as follows. Suppose the Hamiltonian of the
theory is given as the integral of the stress-tensor over a spatial
slice,
$H=\int_{\S} d\S^{\mu}T_{0\mu}$, where as usual time  derivatives of
the fields
have been eliminated in favor of   canonical momenta $\pi$. Then if
one is to
get
Hamilton's equations of motion one should be able to write $\d H
=\int
d\S^{\mu}(A_{\mu}\d\pi + B_{\mu}\d\phi)$. In general however when the
space-time is not spatially closed there will be a boundary term on
the right
hand side of this equation, so in order to get Hamilton's equations
one should
redefine $H$ by adding a boundary term whose variations will cancel
this extra
term. The resulting expression is the generator of time translations.
In
addition in a  generally
covariant theory, there are constraints which imply that the total
stress
tensor (for matter plus gravity) is (weakly) zero, so that the total
energy of
a solution is given by evaluating just this boundary term. By
Hamilton's
equations
it is indeed conserved. In 3+1 dimensions this boundary is the 2
sphere at
infinity and for asymptotically flat solutions the  corresponding
energy is
just  the ADM energy.
In our two dimensional case the boundary of the 1-space is  the set
of points
$\s=\pm\infty$. To obtain the required expression let us write down
the
Hamiltonian as the space integral of $T_{00}$.

$$\eqalign{H_0=&{1\over 2}\int d\s\left [2(\Pi_X^2-\Pi_Y^2)+{1\over
2}(X'^2-Y'^2)+2\sqrt{N\over 12}Y''-4\l^2e^{\sqrt{12\over
N}(X+Y)}\right ]\cr
&+H^f+H^{b,c},\cr}$$
where $\Pi_X,\Pi_Y,$ are momenta canonically conjugate to $X,Y,$ and
a prime
denotes differentiation with respect to the space-like coordinate
$\s$.
Then we have for the variation,

$$\eqalign{\d H_0=&{1\over 2}\int d\s
[4(\Pi_X\d\Pi_X-\Pi_Y\d\Pi_Y)\cr &-\left
(X''+4\l^2\sqrt{12\over N}e^{\sqrt{12\over N}(X+Y)}\right )\d
X+\left
(Y''-4\l^2\sqrt{12\over N}e^{\sqrt{12\over N}(X+Y)}\right )\d Y]\cr
&+\d H^f+\d H^{b,c}
+{1\over 2}[X'\d X-Y'\d Y+2\sqrt{N\over 12}\d
Y']^{\infty}_{-\infty},\cr}\eqn\var$$

{\it assuming that the  matter and ghost configurations are bounded}.

Thus we need to add a boundary term and redefine the Hamiltonian as

$$H=H_0+H_{\pa},$$

such that $\d H_{\pa}$ cancels the boundary term in \var. $H_{\pa}$
cannot be
defined for arbitrary configurations. But for the space of
configurations which
either vanish  asymptotically or go to an arbitrary solution of the
equations
of motion, this may be defined. This is because the general solution
[\bc,\sda
] to the equations of motion is given by

 $$\eqalign{X=&-\sqrt{12\over
N}(u_+(\s^+)+u_-(\s^-))+\l^2\sqrt{12\over N}
\int^{\s^+}d\s^+e^{g_+(\s^+)}\int^{\s^-}d\s^-e^{g_-(\s^-)}\cr
 &=-Y+\sqrt{N\over 12}(g_++g_-),\cr}\eqn\soln$$
 where $u_{\pm}(\s^{\pm})$ are arbitrary chiral functions to be
determined by
the
constraints and $g_{\pm}(\s^{\pm})$ reflect the arbitrariness in the
choice of
conformal frame.\foot{In Kruskal-like coordinates ($x^{\pm}$ of
reference
[\call ]) the latter are zero while in the asymptotically Minkowski
coordinates
 ($\s =\pm{1\over\l}\ln (\pm\l x^{\pm})$) $g_{\pm}=\pm\l\s^{\pm}$.
These
coordinates were called $\hat\s$ in reference [\sda, \shda] while the
Kruskal
coordinates were called  $\s$. In this paper we stick to the notation
of
[\call].} For this space of configurations and variations within this
space, we
have,

$$H_{\pa}=-\sqrt{N\over 12}[{1\over 2}g'(\s )X-X']_{\pa},\eqn\hbdy$$

where $g=g_++g_-$.
 Now since $H_0=0$ (weakly) is a constraint of the theory the energy
is
entirely given by the boundary term. We should however measure energy
relative
to the linear dilaton vacuum (LDV) which corresponds to the solution
with
$u_{\pm}=0$ in \soln. Defining $\D X =X-X_0$ where $X_0$ is the LDV
solution,
we have our final expression for the ADM energy,

$$ E_{ADM}=-\sqrt{N\over 12}[{1\over 2}g'(\s )\D X-\D
X']_{-\infty}^{\infty}\eqn\adm$$

 It is instructive to express this in terms of the original variables
of CGHS
 using \ycdt\ and dropping all quantum corrections i.e. $O(N
e^{2\phi})$ terms.
 Then we get,

 $$E_{ADM}=\D [e^{-2\phi}(\l+2\phi ']_{-\infty}^{\infty}\eqn\admcl$$

At this point it is incumbent upon us to discuss in what coordinate
system this
expression is expected to be valid. As we observed after \var\ the
above
results are valid provided that the matter and ghost configurations
are
bounded. In the quantum theory we cannot make this assumption in
every
conformal frame. The reason is that the stress tensors of matter,
dilaton
gravity and ghosts are not tensors separately due to quantum
anomalies
(Schwartz derivative terms).\foot{The point is that the normal
ordering that is
necessary to define the stress tensor is frame dependent.} It is only
the
total stress tensor that is a tensor and it is zero in every
frame.{\it Thus we
take the point of view that the expressions for the quantum ADM and
Bondi
masses are correct only  in asymptotically Minkowski coordinates.}

Let us evaluate \adm\ for the static quantum solution [\bc , \sda]

$$X=-\sqrt{12\over N}{M_0\over\l}+X_0\eqn\static$$

where $M_0$ is a constant, and $X_0$ is the quantum solution
corresponding to
the linear dilaton vacuum of
the classical theory which is given by

$$\eqalign{X_0=&-\sqrt{12\over N}\left (e^{\l(\s^+-\s^-)}-{ N\over
24}\l(\s^+-\s^-)\right ).\cr}\eqn\ldv$$

Thus in the formula for the ADM energy \adm, we must put  $\D X
=-\sqrt{12\over
N}{M_0\over\l}$. the result is

$$E_{ADM}=0$$

It should be noted that this result is true even in the classical
limit and is
due to the contribution from the negative end of the space.
This is of course consistent with   positive energy theorems since
this
configuration is obtained in the situation where the expectation
value of the
matter stress tensor is zero. In the earlier computation of the ADM
energy
[\sda, \gs ] (following [\ew, \call ]
only the contribution from the $\s\rightarrow\infty$ was kept.
However the
time translation generator that one gets from the Regge Teitelboim
argument
requires the evaluation of the integral at both ends $\s =\pm\infty$.
In
addition the expression \admcl\ differs from the expressions given in
[\ew ]
and [\call ]. In fact this can be calculated directly from the
classical action
using the Regge-Teitelboim argument to get exactly the same result as
\admcl.
The original expression  actually does not give a well defined answer
for the
ADM mass of the static black hole since the solution does not go to
the LDV as
$\s\rightarrow -\infty$.

The apparent existence of negative mass static solutions was pointed
out as  a
problem for the Liouville-like theory [\gs ] since these solutions
are
non-singular (unlike the corresponding static solutions which had
time-like
singularities, and hence could be ruled out on physical grounds). Our
result
shows however that the correct mass for such configurations is always
zero.

To get non-zero mass one has to put in matter. For simplicity  of
presentation
we will explicitly consider only the case of an incoming shock wave
of
$f$-matter $T^f_{++}={ae^{\l\s^+_0}\over\l}\d(\s^+-\s^+_0),~
T^f_{--}=0$ [\call
], but the generalization to  arbitrary  bounded configurations is
trivial.
The conformal frame in which this solution is asymptotically
Minkowski is
related to  is related to the $\s$ frame by
$\ov\s^+=\s^+,~\ov\s^-=-{1\over\l}\ln (e^{-\l\s^-}-{a\over\l})$.

Then the solution is (see for example [\shda ])

 $$\eqalign{-\sqrt{N\over 12}X =& {M\over\l}+e^{\l
(\bar\s^+-\bar\s^-)}-{
N\over 24}\log \left
({e^{\l\bar\s^+}\over\l}\left
({e^{-\l\bar\s^-}\over\l}+{a\over\l^2})\right
)\right )~~~~~for~~\s^+>\s^+_0 \cr
  =& e^{\l\ov\s^+}(e^{-\l\ov\s^-}+{a\over\l^2})-{ N\over 24}\log
\left
({e^{\l\ov\s^+}\over\l}\left
({e^{-\l\ov\s^-}\over\l}+{a\over\l^2})\right
)\right ),~~for~~\s^+<\s^+_0\cr}\eqn\collapse$$

In the above we have put $M=ae^{\l\s^+_0}$ the mass of the classical
black
hole.
In evaluating the ADM mass this needs to be compared with the LDV
solution in
the same  conformal frame i.e. \ldv\ with $\s$ replaced by $\ov\s$.
Then we get

$$E_{ADM}=M+{N\l\over 24}\ln\left (1+{a\over\l}e^{\l (\ov\tau -\ov\s
)}\right
)|_{\ov\s\rightarrow -\infty} +{N\over 24}\l.\eqn\dynadm$$

 Thus  the  quantum anomaly in  the dynamical solution gives an
infinite
contribution to the ADM mass from the  negative infinite end of the
space-like
line. It should be noted  that the classical solution has a well
defined mass
equal to $M$ in these coordinates. The source of the infinite
radiation bath is
quantum mechanical ($N\rightarrow N\hbar$ in the above). It comes
from the fact
that due to the quantum anomaly the solution in the region
$\ov\s^+<\ov\s^+_0$
does not go to the LDV as $\ov\s\rightarrow -\infty$ once we insist
that the
solution for $\s^+>\s^+_0$ is asymptotically Minkowski.
It should be noted that the energy though infinite is positive.

 If we want to represent the situation corresponding to the absence
of a
radiation bath we have to put a boundary as done in [\rst]. We should
stress
here that to do this we need not necessarily use the RST choice of
functions
$h,~ \ov h$ in \pee. The physics may be discussed entirely in terms
of the $X$
and $Y$ fields of the Liouville-like model and the difference between
the two
types of models lies in whether one chooses to put a boundary or not.

The RST boundary condition puts $\pa_{\pm}X=f=0, $ on a critical
curve $X=X_c$,
which
is regarded as the left  boundary of space-time, wherever it is
time-like. Let
us first consider the static solutions \static, \ldv. The RST
boundary is at
$X=X_c=-\sqrt{12\over N}({N\over 24}-{N\over 24}\ln{N\over 24})$. For
$M_0=0$,
i.e. for the LDV, this is the line $\s={1\over 2\l}\ln{N\over 24}$.
For $M_0>0$
on the other hand there is no solution to the boundary curve equation
and it is
not clear how to evaluate $E_{ADM}$. On the other hand for $M_0<0$
there is a
boundary, and since from the boundary conditions there will be no
contribution
to the ADM mass from this end of the space one gets

$$E_{ADM}=M_0.\eqn\admrst$$

 In other words there is no positivity in the theory with boundary.
This is of
course consistent with the fact that there is a negative energy
thunderpop
[\rst] in the dynamical solutions of this theory and as we shall see
later, the
 Bondi mass also reflects this phenomenon.

Let us now consider the collapse scenario. From the explicit solution
it may be
seen that the boundary  curve $X=X_c$ is time-like for $\s^+<\s^+_0$.
Hence the
boundary condition may be imposed there, and in particular it follows
that
there is no contribution to the ADM mass at the left end of the
space. It
should be stressed that this is the case since the boundary is given
by a fixed
value of $X$ (at which $\pa_{\pm}X=0$)  and thus its contribution
cancels
between the collapse solution and the LDV. Thus with the RST boundary
conditions,

$$E_{ADM}=ae^{\l\s^+_0}\equiv M\eqn\mass$$

 If incoming energy is positive ($a>0$) then the ADM mass is
positive. Clearly
this generalizes easily to arbitrary bounded configurations of matter
so that
for these dynamical solutions we have explicitly demonstrated  the
positive
energy theorem for quantum dilaton gravity. It should be pointed out
that in
 deriving the ADM mass for the theory with RST boundary conditions,
we need
only the conditions at the time-like boundary for $\s^+<\s_0^+$. In
order to
obtain the complete physical picture one needs to impose boundary
conditions
 on the critical curve when it becomes time-like again somewhere
above the line
$\s^+=\s^+_0$. As explained by Russo et al [\rst] this leads to a
thunderpop
 - a burst of negative energy to $\cal I^+_L$. We will rediscover
this effect
 when we calculate the Bondi mass.

Let us now discuss the issue of Hawking radiation. As discussed at
length in
section 5 of [\shda ] the usual calculation of Hawking radiation
[\call ]
cannot really be justified in a situation in which one takes back
reaction and
the constraints of general covariance into account. Since this point
does not
seem to be widely appreciated it is perhaps necessary to reiterate it
here. The
constraints tell us that  (the expectation value of) the total stress
tensor is
zero; i.e.
 $$T^{X,Y}_{\pm\pm}+T^f_{\pm\pm}+T^{b.c}_{\pm\pm}=0\eqn\cons$$
 where the last term is the contribution of the reparametrization
ghosts,
together with the equation of motion for the conformal factor
$T_{+-}=0$. The
usual arguments are equivalent to the following. Since only
$f$-matter is
propagating only the $f$ stress tensor should contribute to
 radiation. Now we have the following anomalous transformation law
for the
stress tensor, from the $\s$ coordinate system which covers the whole
space, to
the $\ov\s$ system which covers only the region outside the horizon.

 $$\ov T_{--}^f(\ov\s^-)=T_{--}^f(\s )+{N\over 24}\l^2\left
(1-{1\over
(1+{a\over\l}e^{\l\ov\s})^2}\right )$$
 where the second term on the RHS is the Schwartz derivative for the
transformation from $\s$ to $\ov\s$.
 Then it is argued that $\s^-$ is the appropriate coordinate on $\cal
I^-_R$
and since there should be no incoming radiation on this null line one
must set
 $ T_{--}^f(\s )=0$ so  the Hawking radiation observed on $\cal
I^+_L$
 is just given by the  Schwartz derivative term. However this
argument ignores
the constraint equation \cons. In fact the counting of propagating
degrees of
freedom merely tells us that there are N of them, but this does not
imply that
the energy propagated is  given just by the $f$ stress tensor. In
fact the
physical propagating state must be dressed by $X, Y$ fields
 and possibly ghost fields as well. For the above result to be
compatible with
the constraints there has to be an inflow of $X,Y,b.c$ field energy
to
compensate for the outflow of $f$ energy.

 Thus we believe [\shda ] that the evaluation of the Hawking
radiation must be
done   by calculating the Bondi mass of the system.  This quantity,
like the
ADM mass, could be non-zero for open systems even though the
expectation value
of the total stress tensor is zero. Since  it must give us the total
energy
minus the energy that has been radiated away up to a given retarded
time, it
must be evaluated on a line which is asymptotic to $\ov\s^-=const.$
at
$\s^+\rightarrow\infty$ (i.e. on ${\cal I^+_R}$) and to
 $\ov\s^+=\ov\s^+_1<\ov\s^+_0$ on $\s^-\rightarrow\infty$ (i.e. on
${\cal
I^+_L}$).\foot{One may also take the line to be asymptotic to a
space like
line at the $\ov\s\rightarrow -\infty$ end.} Then, in analogy with
the
expression \adm\ for the the ADM mass, we have for the Bondi mass

$$ E_{Bondi}(\ov\s)=\sqrt{N\over 12}\left [-\l\D X+(\pa_+\D X-\pa_-\D
X)\right
]_{\ov\s^-=\infty ,\ov\s^+=\ov\s^+_1}^{\ov\s^+=+\infty}\eqn\bondi$$

We have to substitute in the above the difference between the
dynamical
solution evaluated in the $\ov\s$ frame and the LDV  solution. Then
we get

$$\eqalign{ E_{Bondi}(\ov\s )=&M_0-{N\over 24}\ln
(1+{a\over\l}e^{\l\ov\s^-})-{N\over 24}{\l\over (1+{\l\over
a}e^{-\l\ov\s^-})}\cr &-\left [-{ N\over 24}\ln
(1+{a\over\l}e^{\l\ov\s^-})-{N\over 24}{\l\over (1+{\l\over
a}e^{-\l\ov\s^-})}\right
]_{\ov\s^-\rightarrow\infty}\cr}\eqn\bondisoln$$

This gives an infinite Bondi mass for any finite value of $\ov\s^-$,
but this
is a reflection of the fact that the ADM mass is infinite. On the
other hand in
the limit $\ov\s^-\rightarrow\infty$ the Bondi mass becomes equal to
$M_0$ i.e.
the initial incoming matter energy.  In [\shda ] by contrast the
Bondi mass was
identified (just as with the ADM) mass as the value evaluated at one
end i.e.
$\cal I_R^+$. However that led to the result that
the Hawking radiation did not stop and the Bondi mass decayed
indefinitely. In
view of our discussion of the ADM mass,  we believe now that the
problem was
the fact that both ends  of the line (space-like in the case of ADM,
light-like
in the Bondi case) need to be considered. This is the correct time
translation
generator in the ADM case and the object that satisfies  positivity.

We have shown that quantum CGHS theory is soluble and satisfies
positivity of
energy. Unfortunately it is not possible to get much insight into
four
dimensional black hole physics from it, since the conformal
invariance forces
us to a
situation where one has to start with an infinite bath of radiation.
It is
clear that to make contact with four dimensional physics one should
have a
time like boundary
(corresponding to the origin of polar coordinates in 3+1 dimensions)
as is done
in the models with the so-called quantum singularity [\banks, \rrst].
However,
in that case  it is not clear whether conformal invariance, which is
a
consequence of general covariance, can be preserved.  Within the
strict
interpretation of the formalism of two dimensional quantum dilaton
gravity, the
theory without boundary seems to be the only consistent one.

However for phenomenological purposes and to check that \bondi\ gives
the
physical picture that one expects in the Hawking evaporation of black
holes let
us evaluate the Bondi mass with RST boundary conditions. In this case
the lower
limit in \bondi\ is replaced by any point on the critical curve
$X=X_c$ for
 $\ov\s^+<\ov\s^+_0$. Then there is no contribution from this end to
the
energy. At the upper end however there are now two regions to
consider. Calling
the point
 where the apparent horizon ($\pa_+X=0$) and the critical curve
intersect
$(\ov\s^+_s,\ov\s^-_s)$, we have the region
 $\ov\s^-<\ov\s^-_s$  (region I of RST; see [\rst] for figure) and
the region
between the time like boundary
 and $\ov\s^-=\ov\s^-_s$ (region II of RST). In region II the black
hole has
decayed and the solution is taken to be the LDV. In region I the
solution is
the
 collapse solution for $\ov\s^+>\ov\s^+_0$. Thus we have

$$\eqalign{-\sqrt{N\over 12}\D X=& 0~~ in ~{\rm II},\cr
=& {M\over\l}-{N\over 24}\ln (1+{a\over\l}e^{\l\ov\s^-})~~in~{\rm
I}.\cr}$$

Computing the Bondi mass from \bondi\ we have,

$$\eqalign{E_{Bondi}(\s^-)=&M_0-{N\over 4}\l\ln
(1+{a\over\l}e^{\l\ov\s^-})-
{N\over 4}{\l\over 1+{\l\over a}e^{-\l\ov\s^-}}~~in ~{\rm I},\cr
=&0~~in~ {\rm II}.\cr}\eqn\ebondi$$

Thus we seem to have a physical picture of Hawking radiation when the
RST
boundary conditions are imposed. For $\ov\s^-\rightarrow -\infty$,
$E_{Bondi}\rightarrow M$; while at late retarded times
$E_{Bondi}\rightarrow
0$. However there is an unphysical feature in the model in that some
time
before $\ov\s^-=\ov\s^-_s=-\ln [{a\over \l}(1-e^{-4M\over\l N})]$ the
Bondi
energy goes negative. Indeed the energy flow is discontinuous at
$\ov\s^-=\ov\s^-_s$.

$$\eqalign{E_{Bondi}(\ov\s_s^-+0)=&0\cr
E_{Bondi}(\ov\s_s^--0)=&-{N\over 4}
(1-e^{-4M\over\l N})\cr}$$

This is just  the effect of the thunderpop of RST [\rst] and is
caused by the
fact that
when the collapse solution is matched to the LDV along the null line
$\ov\s^-=\ov\s^-_s$, the result is  continuous but not smooth.

In conclusion, what we have shown is that in the absence of  a
boundary in the
two dimensional space one is forced to an interpretation where the
black hole
is immersed in an infinite bath of quantum radiation. However
although both the
ADM and Bondi masses are infinite (the latter actually goes to the
collapsing
mass at infinite time) they are positive. On the other hand if RST
boundary
conditions are
imposed one has a physical picture of black hole evaporation at the
price
however of a negative energy thunderpop. The latter is consistent
with the fact
that there are negative energy static solutions in the theory.

{\bf Acknowledgements:}  While this paper was being prepared for
publication
the author had a discussion on the present work with A. Bilal and I.
Kogan,
during which they showed me a paper [\bk] by themselves in which
formula
(9)  is derived in a somewhat different manner. This work is
partially
supported by Department of Energy contract No.
DE-FG02-91-ER-40672.

\refout
\end